	\documentstyle[twocolumn,prl,aps,epsfig,floats]{revtex}
\begin{document}
\draft

\twocolumn[\hsize\textwidth\columnwidth\hsize\csname @twocolumnfalse\endcsname

\newcommand{\ibf}{\mbox{\boldmath $f$}}
\title{Theory of "ferrisuperconductivity" in ${\rm U}_{1-x}{\rm Th}_x{\rm Be}_{13}$
}

\author{
V. Martisovits$^1$, G. Zar\'and$^{1,2}$, and D. L. Cox$^1$
}

\address{
$^1$Department of Physics, University of California Davis, CA 95616\\ 
$^2$Research Group of the Hungarian Academy of Sciences, Institute of Physics,
TU Budapest, H-1521
}
\date{\today}
\maketitle

\begin{abstract}
We construct a two component Ginzburg-Landau theory with coherent pair
motion and incoherent quasiparticles for the 
phase  diagram of ${\rm U}_{1-x}{\rm Th}_x {\rm Be}_{13}$. 
The two staggered superconducting states
live at the Brillouin zone center and the zone boundary, and
coexist for temperatures 
$T\le T_{c2}$ at concentrations 
$x_{c1}\approx 0.02\le x \le x_{c2}\approx 0.04$. We predict 
below $T_{c2}$ appearance of 
a charge density wave (CDW) and Be-sublattice distortion. 
The distortion explains the $\mu$SR relaxation anomaly, 
and Th-impurity mediated  scattering of ultrasound to
CDW fluctuations explains the attenuation peak.
\end{abstract}


]
\narrowtext


{\it Introduction:}  Heavy fermion materials continue to receive intense
experimental and theoretical interest\cite{HFreviews}.  
In particular, the rather spectacular properties of the heavy
fermion superconductor alloy U$_{1-x}$Th$_x$Be$_{13}$ have defied
a comprehensive theoretical understanding for over 15 years.   In this
system, the temperature ($T$) vs. concentration ($x$) plane displays four
distinct ordered phases, as shown schematically in 
Fig. 1\cite{HFreviews,Heffner,Steglich}.   
Pressure studies\cite{Lambert,Rena}, lower
critical field data\cite{Heffner},  and the
size of the specific heat anomalies\cite{HFreviews} 
strongly suggest that all of these
phases are associated with superconductivity.  
However, below $T_{c2}$, local
magnetic probe experiments\cite{Heffner} reveal an anomaly presumed 
to arise from weak magnetism with ordered moment $\mu \le 0.01\mu_B$. 
Longitudinal ultrasound attenuation ($\alpha_s$)
data display a peak, two orders of magnitude above the
normal state background, which contrasts with an $\alpha_s$ {\it drop} in 
standard superconductors\cite{Batlogg}.   
The normal state of this system is also anomalous.
The resistivity of UBe$_{13}$ is reproducibly 
large (order 100 $\mu\Omega$-cm) at
$T_c$ and non-Fermi liquid (NFL) like in its temperature
dependence\cite{HFreviews}(a),\cite{Aronson}. Fermi liquid behavior and a small resistivity are restored
by application of magnetic field and pressure, suggesting that
this residual resistivity is {\it intrinsic}\cite{Aronson}.  
These results
legislate against an interpretation of the
superconductivity of UBe$_{13}$ in terms of pairing within a traditional
Fermi liquid quasiparticle background.  Rather, there is 
electron pair coherence, but electron quasiparticle {\it incoherence}, 
similar to the interlayer
tunneling and ``stripe''\cite{interlayer/stripes} theories of 
cuprate superconductivity.

In this paper, we propose a new phenomenological theory for
superconductivity in U$_{1-x}$Th$_x$Be$_{13}$ which 
explains many of the observed data and also is consistent with the
NFL normal state.  Motivated by results for the two-channel
Kondo impurity\cite{HFreviews}(a),\cite{EK} and
lattice\cite{2ChNormal,JarPangCox} models, we develop a phenomenology in
which localized and presumed odd-frequency pairing excitations hop
coherently on the lattice, but with negative pair hopping/Josephson
coupling from site to site. Our main result is that the low
temperature phase below
$T_{c2}$ is a {\it ferrisuperconductor}, with spatially modulated 
pair amplitude 
and coexisting superconducting order
parameters with zone center and zone boundary center of mass momentum.
The name derives from obvious analogies to ferrimagnetism. 
Moreover, the {\it ferrisuperconductivity} induces
concomitant charge density wave (CDW) and volume conserving 
antiferrodistortive (AFD) order. 
The combined ordering allows us to quantitatively explain 
the anomalies observed in $\mu$SR relaxation\cite{Heffner} and
$\alpha_s$\cite{Batlogg} experiments while invoking {\it no}
magnetic moments in the low temperature phase. For $\alpha_s$, 
our phenomenology is constrained by experiment and 
microscopic theory, leaving one adjustable parameter to provide an
excellent fit to both the temperature and frequency dependence. 
Our theory has several testable predictions. 

{\em Microscopic motivations}:  The two-channel Kondo (Anderson) lattice
model, which has been rigorously solved using dynamical mean field
theory in infinite spatial
dimensions\cite{HFreviews,2ChNormal,JarPangCox}, 
has been shown to display many properties consistent with UBe$_{13}$.
This model consists of local spin 1/2 moments coupled
antiferromagnetically to two species of conduction electrons, 
derived here from the quadrupolar Kondo
effect\cite{HFreviews}(a).  This model naturally explains the NFL
thermodynamics and transport properties of the normal phase. 
The calculations also confirm the possibility of 
an odd-frequency superconducting state\cite{JarPangCox}, as
anticipated from the strong local pairing fluctuations in the impurity
model\cite{HFreviews}(a),\cite{EK}.  The relevant excitations of the
normal state are not conventional Landau quasiparticles,
so coherent single particle transport is
irrelevant.


\begin{figure}
\begin{center}
\psfig{figure=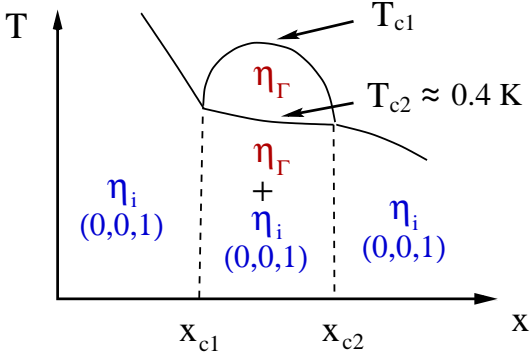,width=7cm}
\end{center}
\vspace*{0.05cm}
\caption{
Sketch of the different superconducting regions and the corresponding 
theoretical
order  parameters for U$_{1-x}$Th$_x$Be$_{13}$. Here, $x_{c1}\approx 0.02$ and 
$x_{c2}\approx 0.04$. The maximum of  $T_{c1}$  is 
$T_{c1}^{max} \approx 0.6\;{\rm K}$. 
$\alpha_s$ peaks at $T_{c2}$, and the muon relaxation rate increases below $T_{c2}$.
In our theory, for  
 $T<T_{c2}$ a CDW and a lattice  distortion 
are generated, which account 
for these data.}
\end{figure}
{\it Ginzburg-Landau Theory:} To characterize  the various superconducting states
we construct a Ginzburg-Landau functional.  
Assuming the superconductivity is induced by 
two-channel Kondo  effect induced pairing interactions at the U 
sites\cite{EK},  one can argue  that the resulting order parameters 
should be odd  in frequency, 
and {\em staggered} in real space\cite{EK,JarPangCox,Piers,staggered}. 
To identify candidate order parameters we investigated a negative pair hopping 
model with third nearest neighbor overlaps, and assumed a 
logarithmically divergent 
local pairing 
susceptibility at the U sites \cite{EK}.  The local
symmetry of these pairing order parameters about the U-sites is
$A_2$\cite{CoxLud}. 
Given the two-sublattice fcc structure of UBe$_{13}$, 
two possible order parameters emerge \cite{third}: 
$\eta_\Gamma$ which sits 
at the Brillouin zone center, 
and the three dimensional order parameter $\eta_i$ which lives at
the three $X$ points $\vec Q_i$
at the  center of the square faces of the fcc Brillouin zone.
These order parameters transform as the odd parity $\Gamma_2^-$ and 
even parity  $X_3^+$ irreps  of the 
$O^6_h(Fm3c)$ space group of ${\rm UBe}_{13}$.  As shown in Fig.
2,  $\eta_\Gamma$ changes in sign from site to
site on the simple cubic U-sublattice, while 
$\eta_i$ has lines of alternating phase parallel to $\vec Q_i$. 

The  resulting free energy functional invariant under time 
reversal, gauge  and  space group operations reads:
\begin{eqnarray}
&&F_0  = \alpha^\Gamma |\eta_\Gamma|^2 + \alpha^X \sum_{i=1}^3 |\eta_i|^2
+ \beta^\Gamma|\eta_\Gamma|^4  + \beta^X_1 \sum_{i=1}^3 |\eta_i|^4\nonumber \\
&& \phantom{F_0} + \beta^X_2 \sum_{i\ne j}^3 |\eta_i|^2|\eta_j|^2
+ \beta^X_3 \sum_{i\ne j}^3 \eta_i^2{\eta_j^*}^2 + F_{\Gamma-X} \;, \label{eq:GL1}\\
&& F_{\Gamma-X} = \xi  \sum_{i=1}^3 |\eta_\Gamma|^2  |\eta_i|^2
+ \zeta  \sum_{i=1}^3 (\eta_\Gamma ^2 {\eta_i^*}^2 + c.c.)\;.
\label{eq:GL2}
\end{eqnarray}
As usual, we assume linear temperature dependence in the 
quadratic coefficients,  {\it viz.},  $\alpha^X = a_1 (T-T_{cX})$ and 
$\alpha^\Gamma = a_2 (T-T_{c\Gamma})$. 

Our conjectured phase diagram is shown in Fig. 1. 
For pure UBe$_{13}$, analysis of observed anisotropies in 
the upper critical
field data\cite{alievold} 
lead us to identify the $\eta_i$-phases as likely
candidates\cite{martobe}.  Pressure experiments
strongly suggest that for $x_{c1}\le x \le x_{c2}$ the lower temperature
phase represents a continuation of the low concentration
line\cite{Lambert,Rena}.  Next, we conjecture that
Th doping produces a relative stabilization of the
$\eta_\Gamma$ phase at intermediate concentrations,  
so that $\eta_i$ and $\eta_{\Gamma}$ coexist for 
$x_{c1}\le x \le x_{c2}$ and $T\le T_{c2}$. 
Given $X$ and $\Gamma$ point order parameter coexistence, we call this
state {\it ferrisuperconductivity} in analogy with ferrimagnetism.  

The parameters, $\beta^X_i$, $\xi$ and $\zeta$ in 
Eqs.~(\ref{eq:GL1}) and (\ref{eq:GL2}) are not known {\em a priori},  
but they are largely   constrained  by the physical 
properties of ${\rm U}_{1-x}{\rm Th}_x{\rm Be}_{13}$. 
Since there is no CDW 
formation, time reversal symmetry breaking, or
splitting of the transition under applied stress in 
pure ${\rm UBe}_{13}$\cite{Rena}, 
the $\beta^X_i$'s must satisfy 
$\beta_2^X>\beta_1^X$  and $\beta^X_3 < \beta^X_2-\beta^X_1$, yielding
a phase with a single $\eta^X_i$ condensed.

The nature of the coexistence phase depends on the sign of 
$\zeta$. For $\zeta>0$  we find a time 
reversal broken coexistence phase, where microscopic current loops 
induce interstitial antiferromagnetic order, a possibility noted
previously for UPt$_3$\cite{Heid2}. 
This could  explain the observed $\mu$SR anomaly. 
However, it is more difficult to explain the 
observed large peak in $\alpha_s$ 
at $T_{c2}$~\cite{Batlogg}. For $\zeta<0$ the mixed superconducting 
state is completely different: in this case a CDW and  a staggered  lattice distortion appear
below $T_{c2}$  and the cubic symmetry of the lattice is lowered to tetragonal
(see Fig.~1).
The CDW arises  because the phasing of the $\eta_\Gamma,\eta_i$ order 
parameters {\it corrugate} the pair amplitude locally, producing a $\pm
|\eta_\Gamma||\eta_i|$ alternation from plane to plane perpendicular to
$\vec Q_i$ (see Fig. 2).  

The anomalies of the second transition can be 
explained consistently and easily for $\zeta<0$: the lattice distortion 
can produce the observed increase in the $\mu$SR relaxation,  
while the peak in $\alpha_s$ results from coupling to large 
induced CDW  fluctuations.  

Henceforth,  we concentrate on the coexistence phase of 
${\rm U}_{1-x}{\rm Th}_x {\rm Be}_{13}$.
To estimate the amplitude of the CDW ($\varrho_i$) and the displacement of the 
face Be atoms ($u_i$) we  first construct  the corresponding symmetry
allowed coupling term in the Ginzburg-Landau
functional:
\begin{eqnarray}
F_{\varrho-\eta-u} &=& \chi_\varrho^{-1}\sum_{i=1}^3 {\varrho_i^2\over2} + 
K_u \sum_{i=1}^3{u_i^2\over2} \nonumber \\
& +&  \gamma \sum_{i=1}^3 \varrho_i u_i 
+ \lambda \sum_{i=1}^3 \varrho_i(\eta_\Gamma \eta_i^* + c.c.) \;,\label{eq:coupling1}
\end{eqnarray}
where the $u_i$'s and the $\varrho_i$'s form three dimensional $X_4^-= \Gamma_2^- \otimes X_3^+$  
$X$ point reps.  The coupling $\gamma\approx 0.4\; {\rm eV/\AA}$ and the 
spring constant $K_u \sim 2 \; 10^4 {\rm K/\AA^2}$ can be obtained from 
microscopic Thomas-Fermi theory calculations and from the bulk modulus, respectively, while $\lambda\approx 10\;{\rm K}$ 
is estimated from the relative shift  of $T_{c2}$ and $\chi_\varrho^{-1} \sim 900\;{\rm K}$ is extracted  
from fits to the  experimental data.

\begin{figure}
\begin{center}
\psfig{figure=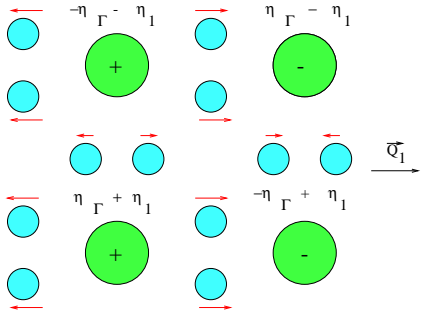,width=5.5cm}
\end{center}
\vspace*{0.05cm}
\caption{Schematic depiction of proposed ordering in U$_{1-x}$Th$_x$Be$_{13}$
below $T_{c2}$.  Large circles are U atoms, small circles represent face
Be atoms, viewed on edge.  $\vec Q_1$ represents the propagation
direction for the assumed $X$ point pair, charge, and displacement
order.  The relative sign of the CDW order is shown by $\pm$
notation on the U atoms.  Arrows represent associated 
displacements of the Be atoms.  }
\end{figure}
As we discuss below, 
from the $\mu$SR experiments we estimate a $T=0K$ lattice
displacement amplitude $u_0= 0.4\;{\rm \AA}$ for Be on planes
perpendicular to $\vec Q_i$.  
Using Eq.~(\ref{eq:coupling1}) a simple calculation yields 
an associated CDW amplitude of  $\varrho_i\le 1.5$, where we
simplistically use $u_0$ in the spring energy to
produce an upper bound. 
This CDW amplitude is quite large at first sight. However,  
the charge is distributed among  the 24 face Be sites surrounding the U ion,
resulting in an excess  charge of $\sim 0.1 \;e/Be$.  
$u_0$ is somewhat large, but in view of the large lattice constant 
$a\approx10\;{\rm \AA}$, is in reasonable agreement with observed 
displacements in  CDW systems\cite{Gruner}. Moreover, it is 
{\em staggered} and volume conserving (to first order). 

The displacement {\em does} result in an 
increased $\mu$SR dephasing rate below $T_{c2}$, with reasoning as 
follows:
The muon sits in the 
center of the Be(II) plaquettes\cite{HeffPriv}.  
Be plaquettes on faces parallel to $\vec Q_i$ are
alternately compressed and expanded.  The dominant
spin-dephasing mechanism of the muons is due to the dipolar coupling
to Be nuclei, and varies in quadrature as $R^{-6}$,
where $R$ is the muon-nearest-Be separation.  Thus, 
muons in expanded(compressed) plaquettes will experience 
reduced(enhanced) relaxation. 
Averaging over inequivalent stopping sites, leaves only
even powers of $u_i$, 
and coupling to compressed plaquettes dominates leading
to an {\em increase} in dephasing.  
The resulting change in the muon  relaxation rate $\sigma_{KT}$ 
is given, to leading order in $u_0$, by 
\begin{equation}
{\delta\sigma_{KT}^2\over \sigma_{KT}^2}  \approx B\; u_0^2 = B \; {4 \lambda^2\gamma^2
\over (\chi_\varrho^{-1} K - \gamma^2)^2}  |\eta_\Gamma|^2|\eta_i|^2\;.
\end{equation} 
with $B \approx 0.4{\rm \AA}^{-2}$. Taking $u_0 =0.5{\rm \AA}$ 
yields 
the estimated zero temperature relaxation rate increase 
of $\sim 10$\%; extending Eq. (4) to all orders in $u_0$ gives 
$u_0\approx$ 0.4${\rm \AA}$ = 0.04$a$, in rough agreement
with NMR imposed constraints\cite{MacLaughlin}.    
The leading order scaling  
with $\eta_i^2$ is in excellent agreement with the 
experimental observations\cite{Heffner}.

Now we analyze the anomalous peak in $\alpha_s$.
As shown in Ref.~\onlinecite{Batlogg}, $\alpha_s$ 
is practically the same for (100) and (111) propagation
directions, 
ruling out the domain wall dissipation mechanism of
Refs.~\onlinecite{Joynt,Varma}. 
The original experimental 
reference\cite{Batlogg} (and recent interpretation of thermal
expansion data\cite{Steglich}) suggest 
that the transition below $T_{c2}$ is purely 
magnetic, citing large moment Gd and Mn
based metallic and insulating magnets with
$\alpha_{s,mag} \approx$10 dB/$\mu$sec at these frequencies.  
However, for magnetic coupling, the two primary contributions to
attenuation are from energy relaxation (which varies like
$\mu^2$), and order parameter
fluctuations (which varies as $\mu^3$)\cite{Kawasaki}. 
These mechanisms are incorrect in temperature dependence and 
magnitude:
given $\mu \le$0.01$\mu_B$ 
from $\mu$SR\cite{Heffner}, we find $\alpha_s \le (.01
\mu_B/.5\mu_B)^2\alpha_{s,mag}  = 4 \times 10^{-5} {\rm
dB}/\mu{\rm sec}$, four orders of magnitude too small.

\begin{figure}
\begin{center}
Fig.a\vskip0.1truecm
\psfig{figure=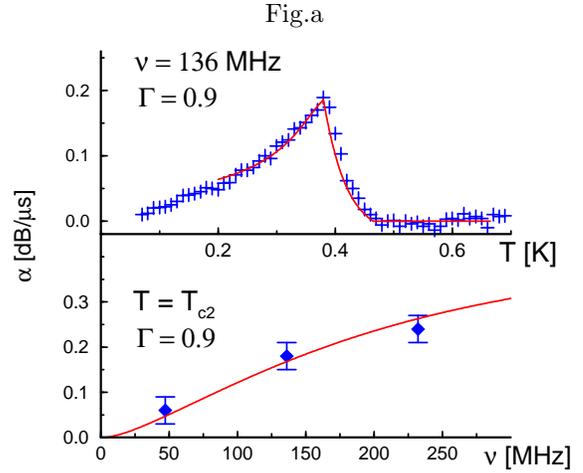,width=7.5cm}
\vskip0.2truecm
Fig.b
\vskip0.4truecm
\psfig{figure=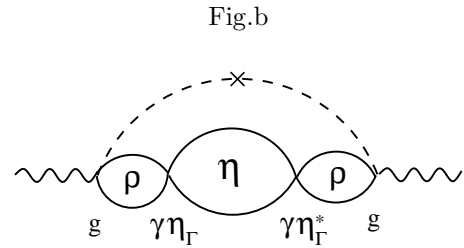,width=6cm}
\end{center}
\vspace*{0.05cm}
\caption{
Fig.~a: Fit to the experimentally observed ultrasound attenuation peak. To 
evaluate the  integral in Eq.~(\protect{\ref{eq:atten}}) we used a cutoff 
$E_c\sim 0.3\;{\rm K}$. The transition temperatures are $T_{c1} = 0.47 \;{\rm K}$
and $T_{c2} = 0.37 \;{\rm K}$. Fig.~b: Leading correction to the ultrasound 
attenuation. The wavy lines indicate the acoustic phonon propagators, while 
the dashed line with the cross stands for impurity scattering at the 
Th sites. Bubbles represent CDW and pairing susceptibilities.}
\end{figure}

In view of these results, and the above interpretation of the $\mu$SR
anomaly, we propose that the $\alpha_s$ peak is due to 
the strong interaction of the impressed phonons 
with CDW fluctuations  at the Th sites. 
Sound waves propagate with a momentum $\vec  q \approx 0$, and therefore  
can linearly couple to the CDW fluctuations at the {\em X} points of the Brillouin zone 
only via the Th impurities at sites $\{\vec R_{imp}\}$. The  
strongest coupling mechanism we find 
results from the destruction of the superconductivity and 
thus the CDW at the non-magnetic Th 
sites, with coupling 
\begin{equation}
\Delta H = g \sum_{\vec R_{\rm imp}} \sum_{i=1}^3  \vec\bigtriangledown \cdot \vec u
(\vec R_{\rm imp}) \;\varrho_i(\vec R_{\rm imp})\;,
\end{equation}
where $\vec u$ denotes the acoustic displacement field and 
$g\approx 0.8\; {\rm  eV}$ is estimated from microscopic model calculations. 
$\alpha_s$ is obtained from the imaginary part of the 
acoustic phonon self energy, shown to leading orxder 
in Fig. 3. 
Replacing the CDW bubbles by the renormalized static susceptibility, 
$\tilde\chi_\varrho = \chi_\varrho /(1-\gamma^2 \chi_\varrho /K)$, and assuming a quadratic 
dispersion  of the form 
$\omega({\bf k}) = C |T-T_{c2}| + D_{\perp} k_\perp^2 + D_\| k_\|^2$
with a cutoff energy $E_c\sim 1K$ we can integrate out the internal  momenta
and we arrive at the following expression: 
\begin{eqnarray}
\alpha_s &\approx & {\rm Im} \{\Pi(\omega)\} = 
C x\;g^2 {\lambda^2 |\eta_\Gamma|^2 \tilde\chi_\varrho^2 \over M c_s^2 \; \Gamma}\; 
I(\omega; T)\;,\label{eq:atten1}\\
I &=& {3\over 2} \int_0^1 dx\; \sqrt{x} {4\Gamma^2 \omega^2
\over (|\alpha_x| + E_c x)^2 + 4\Gamma^2 \omega^2}\;,\label{eq:atten}
\end{eqnarray}
with $C$  a number of the order of unity. To derive Eq.~(\ref{eq:atten}) we assumed 
a hydrodynamic damping $1/\tau \approx  \omega \Gamma$ for the  superconducting 
fluctuations. $\Gamma$ is the only undetermined parameter 
in Eq.~(\ref{eq:atten}); all others  are fixed by microscopic 
model estimates or experimental constraints\cite{CDWcaveat}. Substituting the different couplings into 
Eq.~(\ref{eq:atten1}) we estimate the  prefactor to be $20\; \Gamma^{-1} {\rm dB}/\mu{\rm s}$, which is in 
very good agreement with the experiments (away from criticality, the overall
magnitude is considerably reduced by the integral in Eq. (6)). 
Note that: 
(i) $\alpha_s$ is asymmetric about the transition due to the 
$\eta_\Gamma^2$ factor; (ii) for $\omega\to 0$, 
$\alpha_s \sim |T-T_{c2}|^{-1/2}$ away from $T_{c2}$; (iii) for
$T=T_{c2}$ and $\omega \to 0$, $\alpha_s \sim \omega^{3/2}$.  
We obtain a fit of remarkable quality to the $\alpha_s$ data of 
Ref.~\cite{Batlogg}, as shown  in  Fig.~3.

{\em Predictions and further considerations:}  
Our theory admits a number
of falsifiable predictions below $T_{c2}$:
{\em (1)} The Be displacements with magnitude
$\le 0.4 {\rm \AA}$ 
should be easily observable in neutron scattering experiments.
{\em (2)} The impurity scattering mechanism produces an isotropic angular
dependence for $\alpha_s$
above $T_{c2}$ and zero attenuation for transverse ultrasound,  
consistent with the rough agreement 
between longitudinal attenuation coefficients along
the (100) and (111) directions measured for different samples of
comparable nominal composition\cite{Batlogg}.  We anticipate an order of
magnitude difference between longitudinal and transverse attenuation. 
{\em (3)} The $\alpha_s$ peak will be maximized at the concentration
$x\approx 0.03$ and drop significantly in magnitude near
$x_{c1},x_{c2}$.  {\em Assuming insignificant variation of
electronic parameters as a function of $x$}, then
$\alpha_s(x)/(T_{c1}(x)-T_{c2}(x))$ should vary as $x$
according to Eq. (6).  This may provide a test of the impurity scattering
mechanism. {\em (4)}. The observed NMR line broadening below $T_{c2}$ of
Ref.\cite{MacLaughlin} was originally attributed to the flux lattice,
but this is ruled out by subsequent penetration depth data\cite{london}.
The broadening is roughly consistent with field gradient shifts due 
to our proposed Be sublattice distortion.

{\em Acknowledgements}.  We thank  A. Castro-Neto,
R. H. Heffner, R. Joynt, D.E. MacLaughlin, M.B. Maple, and R. Zieve  for
useful discussions.
This research has been supported by 
the U.S - Hungarian Joint Fund Nr. 587, and grant No. DE-FG03-97ER45640 of the
U.S DOE Office of Science, Division of Materials Research, 
and by Hungarian Grants OTKA~Nrs.~T026327 and F030041.



\begin{thebibliography}{99}
\bibitem{third}
With third nearest neighbor
hopping included, the leading order gradient free energies 
of both order parameters can be made 
locally stable.
\bibitem{CDWcaveat} We assume full reduction of the 
CDW amplitude at Th sites, which produces an upper bound
to $\alpha_s$.
\end{thebibliography}
\end{document}